\documentclass[12pt]{article}
\usepackage[margin=1in]{geometry}
\usepackage[affil-it]{authblk}
\usepackage[utf8]{inputenc} 
\usepackage[T1]{fontenc}    
\usepackage[hidelinks]{hyperref}       
\usepackage{url}            
\usepackage{booktabs}       
\usepackage{amsfonts}       
\usepackage{pifont}
\usepackage{nicefrac}       
\usepackage{microtype}      
\usepackage{lipsum}
\usepackage[]{url} 
\usepackage{hhline, soul}
\usepackage{hyperref}
\usepackage{svg}
\usepackage{tabularx}
\usepackage{framed}

\usepackage[backend=biber,sorting=none]{biblatex}
\usepackage{subcaption}
\usepackage{mathtools}
\usepackage{stackengine}
\setstackEOL{\\}
\usepackage{tikz}
\usepackage{tikz-3dplot}
\usepackage{caption}
\usepackage{tikz-cd}
\usetikzlibrary{arrows,decorations.pathmorphing,backgrounds,positioning,fit,petri,calc,shapes.misc,decorations.markings,intersections}

\usepackage{booktabs}
\usepackage{siunitx}

\usepackage{amssymb}
\usepackage{amsmath,amsfonts,amsthm} 
\usepackage{stmaryrd}
\usepackage{enumitem}
\usepackage{array}
\usepackage{scalerel}

\usepackage{verbatim}
\usepackage{tensor}
\usepackage{enumitem}
\usetikzlibrary{backgrounds}
\usepackage{float}
\usepackage{csquotes}

\title{Community agreements pedagogy in physics}
\author[1]{Rachel E. Scherr}
\author[2]{Jer Steeger}
\author[3]{Paige Sechrest}
\affil[1]{Department of Physics, University of Washington Bothell, Seattle, WA\\ rescherr@uw.edu}
\affil[2]{Department of Philosophy, University of Bristol, Bristol, UK\\ jer.steeger@bristol.ac.uk}
\affil[3]{Political Science and Criminal Justice Department, University of Wisconsin-Eau Claire, Eau Claire, WI\\ sechrepa@uwec.edu}

\date{}

\begin{document}
\maketitle




\tableofcontents

\section{Introduction}

The idea of engaging critically and creatively with physics theories is a stretch for many physicists and physics educators. Most physicists become enculturated into the belief that physics definitions, laws, and equations are read from the book of nature \cite{Traweek_1988,Harding_2015}. However, this belief obscures how theories are collectively negotiated by people in specific socio-political and personal conditions—and how students’ particular ideas and intuitions will inevitably shape their own practice of science. To help students build their skills for collective theory-building, we suggest a pedagogy for physics in which students actively negotiate the laws and definitions of theories.

The instructional format consists of worksheet-based, small-group activities in which students are invited to treat proposals for the laws and definitions of a physical theory as parts of a community agreement (which we call a “physics agreement”). These activities are in the style of Tutorials in Introductory Physics \cite{McDermott_Shaffer_2012}, which lead students step by step to think more deeply about physics concepts, talk to each other, and develop ideas together, with specific learning goals for specific science topics. Unlike those materials, ours invite open-ended creative contributions at key moments as a means of increasing students’ epistemic agency, i.e., their ability to contribute to scientific knowledge. Modeled after practices in equity education \cite{Arao_Clemens_2013,Simon_Boyd_Subica_2022,Developing_community_agreements}, the approach validates students as knowers in a physics community and teaches them skills for collective theory-building. The approach retains more guidance and structure than open-inquiry formats such as Student-Generated Scientific Inquiry, in which students generate their own questions and methods to pursue them \cite{Salter_Atkins_2013}. Illustrations are drawn from an undergraduate course in spacetime physics, but the approach is not specific to this topic. (Spacetime physics lends itself especially well to an agreements approach in that it is based on just a few key assertions with far-reaching implications, but all subject areas in physics have this quality to some degree.) The course we describe took place at a medium-size public university whose Carnegie classification is “higher access,” indicating that it offers broad access to learners, particularly those who have been underserved \cite{2025_Student_Access_and_Earnings_Classification}.

\section{Community agreements}

“Community agreements,” also known as norms or ground rules, are commonly used by facilitators in equity education to support civil discussion of potentially sensitive or divisive topics. Traditionally, they address relational values, rather than class content, by indicating intentions such as “listen to understand” and “critique ideas not people” \cite{Arao_Clemens_2013,Simon_Boyd_Subica_2022,Developing_community_agreements}. Making community agreements explicit and negotiable shares power among social groups and is helpful for discussing challenging topics, which may include strong emotions and vulnerable conversation across social differences. A collectivist approach, in which participants create the agreement themselves, demonstrates this power-sharing most clearly; however, time and size constraints might require a facilitator to present the ground rules themselves, or adopt a hybrid approach where some are suggested and others are added by participants \cite{Arao_Clemens_2013,Simon_Boyd_Subica_2022,Developing_community_agreements}. The course that we will describe below began with negotiating a community agreement that addressed a brief grace period at the start of class, including all learners in discussions, and acceptance of potentially strong emotions in a physics learning context.

\section{Physics agreements}
Our approach extends community agreements to apply to disciplinary content. We invite students to adopt physical laws and definitions as norms that will structure how they complete in-class activities, norms we refer to as items in a “physics agreement.” Key features of a physics agreement include instructor-provided starting points for negotiation, regular student-led revision of the developing theory, official documentation, and continual support for students to make choices about their knowledge-building activities; we describe each of these features below.

Like community agreements, instructors can restrict the scope of physics agreements in line with time and size constraints. Somewhat unlike community agreements, the scope should be in line with content learning goals for students. At one end of the spectrum, the instructor might provide students with a perplexing thought experiment and ask students to come up with all the rules on their own; certain tutorials can be viewed as exploring something like this structure \cite{Robertson_Goodhew_Bauman_Heron}. At the other end, the instructor might invite students to tweak, clarify, or elaborate instructor-provided “proposals” for physics agreements. The examples that follow are from a course in which we took the latter approach. Our preliminary observations suggest that even this fairly structured and minimal redistribution of power has the potential to improve students’ confidence, engagement, and understanding.

\subsection{Provide starting points}
When beginning study of a new topic area in physics, it is common for the instructor to set out initial definitions, laws, or guidelines: for example, the study of kinematics often begins with the instructor providing definitions of displacement and velocity. In my (author RES) spacetime physics class, the first collaborative exercise students did together included instructor-provided definitions of “reference frame” and “event,” two key concepts in spacetime physics. The exercise was a collaborative worksheet in the style of Tutorials in Introductory Physics \cite{McDermott_Shaffer_2012}, but was distinctive in that these initial definitions were displayed in a box on the worksheet titled “Proposed Physics Agreement.” The physics agreement provided a common language amidst differing interpretations, setting shared standards for the theory under development \cite{Longino_1990,Borgerson_2011}. Highlighting these items with a box made it easier for students to refer to them, set a gentle expectation that they would do so, and made it easier to prompt them to do so.

\subsection{Invite adjustment}
The “Proposed Agreement” terminology is meant to convey that the instructor’s suggestions are provisional, subject to change by mutual consent as the project is realized. This revision is essential to the practice of theory development. Thus, I routinely invited students to reflect and synthesize in small groups by ending each tutorial worksheet with an invitation such as: “Is there anything you want to challenge, add, rephrase, or clarify about the agreement so far?” My students responded definitively to this invitation with multiple modifications that felt important to them. For example, the definition of an event proposed by the instructor was “An event is associated with a single location in space and a single instant in time”: one of the additions requested by students was, “Don’t go thinking that when you received the signal is when it happened; those are distinct events,” representing an issue they had struggled with in the tutorial.

The students’ readiness to add to the physics agreement was at least partly due to their experience adding to the community agreement: they already created a supportive learning environment, they understood that the class was their responsibility as well as mine, and they knew that I was serious about taking the substance of their ideas into account \cite{Coffey_Hammer_Levin_Grant_2011}. One student, Parbeen, highlighted how the two agreements worked together to support collective learning:
\begin{quote}
    “The community agreement and physics agreement… makes me look forward to class sessions because I feel like I have a voice in how things are conducted. This sense of involvement and influence is empowering and makes me more enthusiastic about learning. The agreement also fosters a stronger sense of community and mutual respect. By creating a document that everyone has contributed to, we ensure that all voices are heard and considered… It’s comforting to know that everyone is on the same page and that we have collectively agreed on how to handle various aspects of the class.”
\end{quote}

\subsection{Make it official}
As the student above noted, clear documentation of the agreement is key to its functioning as a shared public standard \cite{Longino_1990,Borgerson_2011}. In my class, both the community agreement and the physics agreement were recorded in an editable document that was accessible to all participants and displayed daily in class. I deliberately treated this document as an essential resource, encouraging students to refer to it and explain their reasoning in terms of specific items. I graded assignments and gave feedback not in terms of “correctness,” but in terms of alignment with the physics agreement as we had articulated it up to that point. As the agreement became longer and more elaborate, the students and I created short names for key parts of the agreement (shown in Fig.~\ref{fig:phys-agree} as “Event,” “Reference frame,” etc.); this supported us all to refer to specific items of the agreement in conversation and allowed me to include an abbreviated version of the agreement on later collaborative worksheets. 

\begin{figure}
    \begin{center}
    \setlength\fboxsep{.4cm}
    \fbox{%
      \parbox{.75\textwidth}{
        \textit{Physics agreement}
        \begin{itemize}[leftmargin=*]
            \item \textbf{Reference frame:} An observer's reference frame is an arrangement of assistants and equipment with which the observer may record the position and time of anything that occurs.
            \item \textbf{Event:} An event is associated with a single location in space and a single instant in time.
            \begin{itemize}
                \item Don’t go thinking that when you received the signal is when it happened; those are distinct events.
            \end{itemize}
            \item \textbf{Laws are the same:} All the laws of physics are the same in every free-float frame.
            \item \textbf{Light speed is the same:} The speed of light is the same in all directions in all free-float frames and is the fastest speed possible.
        \end{itemize}
      }%
    }   
    \end{center}
    \caption{Partial example of a physics agreement for spacetime physics.}
    \label{fig:phys-agree}
\end{figure}

\subsection{Normalize revision}
The power of the physics agreement became more strongly evident in my class when the students entered territory that was not covered by the agreement up to that point: in my case, when they encountered the relativity of simultaneity. This encounter provided their first opportunity to not only adjust the agreement, but also substantively revise it, including explicit rebuttal of a physics idea they had formerly (implicitly) agreed upon. 

During the class period in which students encountered the relativity of simultaneity, the physics agreement items included statements of the two relativity postulates, expressed as “laws are the same” (all the laws of physics are the same in every inertial frame of reference) and “light speed is the same” (the speed of light is the same in all directions in all inertial frames and is the fastest speed possible). The pivotal scenario, which we call the “tape player” scenario, is described elsewhere in detail \cite{Scherr_2007,Scherr_Shaffer_Vokos_2002}. The tape player scenario takes place in the context of a modified version of Einstein’s train paradox, in which two sparks occur at either end of a train that moves with relativistic speed relative to an observer Alan who is at rest on the ground. The sparks are simultaneous in Alan’s frame. Another observer, Beth, is standing at the center of the train. Students are asked whether, in Alan’s reference frame, Beth receives the wavefront from the front spark (wavefront $F$) before, after, or at the same time as the wavefront from the rear spark (wavefront $R$). Most students recognize that in Alan’s frame, Beth receives wavefront $F$ before wavefront $R$ because in Alan’s frame she is moving toward the center of the front wavefront. In articulating this result, students explicitly referenced the “light speed is the same” item of the physics agreement.

The next step is for students to encounter a situation that requires them to critically assess and ultimately revise the physics agreement: this takes place as they determine the order of events in Beth’s frame. To assist students with this part of the analysis, we introduce a cassette tape player at Beth’s feet that starts or stops playing music when a wavefront hits it (and stays silent if two wavefronts hit it at the same time). Students are asked whether the tape player plays in Alan’s frame and in Beth’s frame. The analysis in Alan’s frame shows that Beth receives wavefront $F$ before wavefront $R$, and thus the tape player plays. Students do not immediately think that if the tape player plays in any frame, it should play in all frames, as special relativity dictates. Instead, many claim that the music plays in Alan’s frame but not in Beth’s frame, combining the agreement item that “light speed is the same” with the implicit idea that events that are simultaneous in Alan’s frame are also simultaneous in Beth’s frame. Subsequent questions in the tutorial ask whether Beth will hear the music and whether Beth will later observe the tape to have advanced from its starting position, further challenging students’ conclusions.

The end of the tutorial includes reflection questions to help students slow down, acknowledge the disruption they were experiencing, and tune in to their emotional response: e.g., “Are you concerned about any of the conclusions you have reached so far? What are your concerns?” and “What did you feel most prominently when working on those questions, including both physical sensations and emotions?” Finally, the tutorial invites students to revisit the physics agreement as usual, including drafting their own additions.

In the ensuing whole-class discussion, students noted how their difficulty stemmed from something tacit or underspecified in the physics agreement: the idea that time runs the same for everybody. This led to collaborating on revisions and directing me to make certain official changes to the agreement:

\begin{quote}
    \textbf{Joey:} I guess it was just a lot of confusion surrounding the two reference frames because the first physics agreement that we had was working against us in that.
    
    [...]
    
    \textbf{Alex:} So if I asked you, on number two, the physics agreement where it says an event is associated with a single location in space and a single instant in time. Would you say that would work only if it's in a specific reference frame?
    
    \textbf{RES:} You're saying in another reference frame, an event would not be associated with a single instant? Maybe not the same instant.
    
    [...]
    
    \textbf{Corben:} I would still add that though. The difference in which instants and locations they are, depending on which reference frame you're in.
    
    \textbf{Connor:} How much they aren't simultaneous depends on what frame of reference you're looking at it from.
\end{quote}
In response, I wrote “Time and position of an event depends on reference frame” on the board, and added it to the official physics agreement.

In subsequent classes, students continued to add to the physics agreement as they encountered new situations, thereby constructing a substantial and thorough theory of special relativity. One student, Joey, described the process this way in a written reflection:
\begin{quote}
    “Oftentimes we’d have set definitions or specific constraints added to the agreement based on the scenarios in class or in homework that acted as a basis of knowledge we could work around. However, that doesn’t mean that what we are presented with is the end all be all. There were a couple instances, especially with simultaneity, that forced us (the class) to really question whether or not our previously mentioned constraints on a topic were enough or if we needed to elevate our constraints based on the new scenario presented to us. Whether we added something to the agreement later on also had a number of challenges because we always had to consider whether any violations were occurring with new additions, which provided a more and more complicated feedback loop the longer the class goes on.”
\end{quote}

\subsection{Support agency}
The relativity of simultaneity challenges the common belief that, roughly, time runs the same for everyone. This belief is often deeply-held, and its denial can challenge students’ integrity as epistemic agents: research documents learners responding with withdrawal and absurdism \cite{Scherr_2007,Scherr_Shaffer_Vokos_2002}, and as instructors we have unfortunately borne witness to self-abasement (“I’m so dumb”), refusal (“I never want to think about this again”), and capitulation (“whatever you say, boss”). In contrast, students in this class described conditions of psychological safety that supported continuous, collective learning. As one student, Carol, put it:
\begin{quote}
    “The safety of being able to be wrong and then discovering that you were wrong and amending your beliefs…[it’s] like when children learn something fundamental, and they think that that's the truth, and as their learning develops they realize, oh, that's not the universal truth, there are conditions to that. Realizing as you're working through problems, how it's wrong, and being able to tack on conditions that then make it more right is what it felt like–and doing it together. Nobody took the blame for initially being wrong.”
\end{quote}
We propose that the physics agreement cultivates students’ epistemic agency in at least two ways. First, the agreement process externalizes the challenge as an obstacle for a community project, instead of a threat to self. Some agreement items may have consequences that were previously overlooked, but not because an individual student was wrong; rather, recognizing those consequences is part of the collective process of exploring uncharted territory. The puzzle is framed not as a challenge to a student’s integrity as a knower, but as an opportunity to strengthen their membership in the class’s knowledge-making community. Second, the agreement process leaves open multiple options for students’ own beliefs (e.g., about time). They may revise their personal beliefs in line with the physics agreement \cite{Scherr_2007}; they may maintain their personal beliefs, but accept the agreement as if it is true in the context of spacetime physics \cite{Fine_1986}; or they may consider other hypothetical physics agreements that align better with their personal beliefs \cite{Darrigol_2022}. Having a choice about their relationship to the physics agreement deepens their autonomy as knowers.

\section{Summary}
We propose that co-creating a physics theory as a negotiated physics agreement strengthens students’ self-concept as knowers and inducts them into authentic scientific practices. A pedagogy of community agreements and physics agreements is designed to share power with students, increase their investment in the theory, reduce harm, and teach practices of collective theory-building. In the spacetime physics class, students said the process “allows us to build the theory as if we created it ourselves,” and stated that it “promotes a sense of responsibility and accountability. When everyone contributes, we all become more invested in the success of the class, creating a more supportive and cohesive learning environment.”

\section*{Acknowledgements}
The authors are grateful to the students in the spacetime physics class described in this article, including Tammy Chau, Joey Del Gianni, Alexander Kennedy, Carol Miu, Imad Morsli, Parbeen Sekhon, and Connor Wiedmann. Preperation of this manuscript was supported by British Academy Grant No.~IF23\textbackslash100451. This manuscript contains material based upon work supported by National Science Foundation Grant No.~1936601.

\printbibliography

\end{document}